\documentclass[english,twoside,a4paper,10pt]{article}

\usepackage[latin1]{inputenc}
\usepackage[T1]{fontenc}
\usepackage{amsmath}
\usepackage{amsfonts}
\usepackage{graphicx}
\usepackage{subfigure}
\usepackage{a4wide}
\usepackage{amssymb}
\usepackage{fancyhdr}
\usepackage{mathrsfs}
\usepackage[toc,page]{appendix}

\begin{document}

\title{\bf Inhomogeneous fluids for warm inflation}
\author{ 
Shynaray Myrzakul\footnote{Email: shynaray1981@gmail.com},\,\,\,
Ratbay Myrzakulov\footnote{Email: rmyrzakulov@gmail.com},\,\,\,
Lorenzo Sebastiani\footnote{E-mail address: l.sebastiani@science.unitn.it
}\\
\\
\begin{small}
Department of General \& Theoretical Physics and Eurasian Center for
\end{small}\\
\begin{small} 
Theoretical Physics, Eurasian National University, Astana 010008, Kazakhstan
\end{small}\\
}

\date{}

\maketitle

%%%%%%%%%%%%%%%%%%%%%%%%%%%%%%%%%%%%%%%%%%%%%%%%%%%%%%%%%%%%%%%%%%%%%%%%%%%%%%%%%%%%%%%%%%%%%%%%%%%%%%%%%%%%%%%%%%%%%%%%%%%%%%%%%%%

%%%%%%%%%%%%%%%%%%%%%
%  Abstract
%%%%%%%%%%%%%%%%%%%%%
\begin{abstract}
Inhomogeneous fluid models for warm inflation are investigated. The early-time acceleration is supported by
inhomogeneous fluid whose coupling with radiation leads to the radiation dominated era after inflation. Several examples are analyzed, strong dissipation regime is discussed, and the viability of the models respect to the last Planck data is verified. 
\end{abstract}
%%%%%%%%%%%%%%%%%%%%%

%----------------------------
%PACS
%----------------------------

%===========================================================================

\tableofcontents
%%%%%%%%%%%%%%%%%%%%%%%%%%%
%%%  Sec. I
%%%%%%%%%%%%%%%%%%%%%%%%%%%
\section{Introduction}

Inflation was suggested by Guth and Sato~\cite{Guth, Sato} several years ago to solve the problems of initial conditions of Friedmann universe, and today it is well accepted the idea  
according to which the universe underwent a period of strong accelerated expansion after the Big Bang. Despite to the fact that the arena of the models for early-time acceleration is quite large, the constraints that the theory must satisfy to reprouce the last cosmological data~\cite{WMAP, Planckdata} are quite restrictive (see Refs.~\cite{Linde, revinflazione} for general reviews about inflation).

Inflation takes place at the Planck time ($\sim 10^{-35/-36}$ sec.) and
brings to the thermalization of observable universe: since it accelerates the expansion, small initial velocities within a causally connected patch become very large and horizon and flatness issues can be well explained. To produce an acceleration one needs repulsive gravity. All the data indicate that inflation was realized by a (quasi) de Sitter solution, but a mechanism to quikly exit from acceleration is necessary to recover radiation/matter dominated universe.

Chaotic inflation~\cite{chaotic, buca1, buca2, buca3, buca4} is based on a scalar field, dubbed ``inflaton'', subjected to some suitable potential. The inflaton supports acceleration when its magnitude and the potential are large: at the end of inflation, the field falls down in a minimum of the potential and acceleration ends. Scalar field representation is not the only possibility to describe inflation: other models are based on fluid cosmology or on modified theories of gravity (see Refs.~\cite{reviewmod, reviewmod2, Caprev, myrev} for review), like Starobinsky model~\cite{Staro} with the account of a $R^2$-term in the action of General Relativity and which is in agreement with observational data (see also Refs.~\cite{mioStaro, mioricostruzione}).

Since during strong accelerated expansion all the matter/radiation components of the universe are shifted away, after inflation some reheating process for particle production is necessary. In scalar field inflation, when the inflaton reaches the minimum of the potential, it starts to oscillate and, due to a coupling with the particle fields, transfers its energy to matter/radiation.
An alternative to this standard scenario is given by warm inflation~\cite{berera1,berera2, berera3, maeda,bartrum}, where the production of radiation occours during inflation. Since repulsive energy must be dominant and the radiation energy has to be not shifted away, a dissipative term in the equations of motion must be introduced and, at the end of inflation, one recovers the radiation dominated universe without invoking any reheating.

The aim of this paper is to investigate warm inflation in fluid comsology, where a ``dark'' inhomogeneous fluid coupled with radiation leads to accelerated expansion with the production of radiation/ultrarelativistic matter.
General introduction to cosmology of inhomogeneous fluids is given in Refs.~\cite{fluidsOd, Capo, Odi1}.

The paper is organized as follows. In Section {\bf 2}, we will revisit warm inflation in scalar field theories. In section {\bf 3}, we present the formalism of fluid warm inflation. Section {\bf 4} is devoted to the analysis of several examples of warm inflation with inhomogeneous fluids. Different choices of dissipative term are investigated and for every case the viability of the model respect to the last Planck data and the emerging of radiation dominated universe after inflation are analyzed. Conclusions and final remarks are given in Section {\bf 5}.

%%% Unit %%%
We use units of $k_{\mathrm{B}} = c = \hbar = 1$ and denote the
gravitational constant, $G_N$, by $\kappa^2\equiv 8 \pi G_{N}$, such that
$G_{N}^{-1/2} =M_{\mathrm{Pl}}$, $M_{\mathrm{Pl}} =1.2 \times 10^{19}$ GeV being the Planck mass.
%%%%%%%%%%%%
%%%%%

%%%%%%%%%%%%%%%%%%%%%%%%%%%%%%%%%%%%%%%%%%%%%%%%%%%%%%%%%%%%%%%%%%%%%%%%%%%%%%%%%%%%%%%%%%%%%%%%%%%%%%%%%%%%%%%%%%%%%%%%%%%%%%%%%%%

\section{Warm inflation in scalar field theories}

A simple description of warm inflation in the scalar field representation is given by the introduction of a friction term between the scalar field driving inflation (the inflaton) and the ultrarelativistic matter/radiation in primordial universe. 
In the classical approach, matter and radiation are totally negligible (or absent)
during inflation: due to the huge energy density of the inflaton, all the other components of the universe are shifted away, and only after the end of inflation some reheating process for particle production occours. However, one may suppose that, at least at the beginning of inflation, there is thermal contact between inflaton and the radiation fields with which it interacts.
Since the inflaton is almost a constant, this interaction 
remains the same, the thermalized radiation
bath of ultrarelativistic matter and radiation do not vanish with expansion and we do not need reheating after inflation. 

In this section, we would like to briefly review how warm inflation is realized in the presence of a scalar homogeneous field $\phi$ whose Lagrangian is given by
\begin{equation}
\mathcal L_{\phi}= -\frac{\partial^\mu\phi\partial_\mu\phi}{2}-V(\phi)+L_{int}\,,
\end{equation}
where $\phi$ represents the inflaton subjected to the potential $V(\phi)$ and producing acceleration, and $L_{int}$ describes its interaction with radiation fields. On flat Friedmann-Robertson-Walker space-time,
\begin{equation}
ds^2=-dt^2+a(t)^2 d{\bf x}^2\,,\label{metric}
\end{equation}
where $a(t)$ is the scale factor depending on the comsological time $t$, the Friedmann equations read
\begin{equation}
\frac{3 
H^2}{\kappa^2}=\rho_{rad}+\rho_\phi\,,\quad-\frac{1}{\kappa^2}\left(2\dot 
H+3 H^2\right)=p_{rad}+p_\phi\,,
\end{equation}
where $H(t)$=$\dot a(t)/a(t)$ is the Hubble parameter, the dot being the derivative with respect to the time, $\rho_{rad}\,,p_{rad}$ are the energy density and the pressure of radiation and $\rho_\phi$ and $p_\phi$ correspond to the energy density and the pressure of the inflaton,
\begin{equation}
\rho_\phi=\frac{\dot\sigma^2}{2}+V(\sigma)\,,\quad 
p_\phi=\frac{\dot\sigma^2}{2}-V(\sigma)\,.\label{relphi}
\end{equation}
The equation of state of radiation simply reads
\begin{equation}
p_{rad}=\frac{\rho_{rad}}{3}\,.
\end{equation}
The friction term appears in the (total) conservation law of the model and leads to
\begin{eqnarray}
\ddot\phi+3H\dot\phi+V'(\phi)&=&-\mathcal Y\dot\phi\,,\nonumber\\
\dot\rho_{rad}+4H\rho_{rad}&=&\mathcal Y\dot\phi^2\,,\quad\quad 0<\mathcal Y\,.\label{consrad}
\end{eqnarray}
Here, the prime denotes the derivative respect to the field and the friction term $\mathcal Y\dot\phi$ 
comes from 
$\mathcal L_{int}$ and describes the decay of the inflaton. In the standard inflationary scenario, this term becomes important after inflation and it is responsable for the reheating. Here, we have to assume $\mathcal Y\sim 3H$ during inflation, such that we must take into account its contribute. 

Inflation is realized by a (quasi) de Sitter solution, when the energy density of inflaton is dominant ($\rho_{rad}\ll\rho_\phi$) and almost a constant, and its kinetic energy is negligible respect to the potential ($\dot\phi^2/2\ll V(\phi)$).
The magnitude of the (negative) field is assumed to be very large and in the ``slow-roll approximation'' the first Friedmann equation with the first equation in (\ref{consrad}) read
\begin{equation}
\frac{3H^2}{\kappa^2}\simeq V(\phi)\,,\quad\dot\phi\simeq-\frac{V'(\phi)}{\left(3H+\mathcal Y\right)}\,.\label{sa}
\end{equation}
The field slowly grows up during inflation (thus, $0<\mathcal Y\dot\phi$) and therefore the Hubble parameter decreases if $V'(\phi)<0$. 
If we introduce the slow-roll parameters
\begin{equation}
\epsilon=\frac{1}{2\kappa^2}\left(\frac{V'(\phi)}{V(\phi)}\right)^2\,,\quad
\eta=\frac{1}{\kappa^2}\left(\frac{V''(\phi)}{V(\phi)}\right)\,,
\end{equation}
where $\epsilon$ always is positive (while $\eta$ is in general negative),
since for the (quasi) de Sitter solution of inflation it must be $|\dot H/H^2|\,,|\ddot H/(H\dot H)|\ll 1$, one obtains
\begin{equation}
\epsilon\,,|\eta|\ll 1+Q\,,\quad Q=\frac{\mathcal Y}{3H}\,.\label{epsilonetaphi}
\end{equation}
As a consequnce, in the strong dissipation regime $1<Q$, in slow roll approximation one may have $1<\epsilon\,,|\eta|$. In other words, we do not necessarly need ``flat'' potential like in the classical chaotic inflation.

Let us see what happen to radiation. From (\ref{consrad}) we get
\begin{equation}
\rho_{rad}\simeq\frac{\epsilon Q V(\phi)}{2(1+Q)^2}\,,
\end{equation}
and the energy density of radiation is almost a constant during inflation, when we have thermalization and $\rho_{rad}\sim T^4$, $T$ being the temperature of the radiation bath. We immediatly note that
\begin{equation}
\frac{\rho_{rad}}{\rho_{\phi}}\simeq\frac{\epsilon Q }{2(1+Q)^2}\,,
\end{equation}
and $\rho_{rad}$ is negligible when $\epsilon\ll 1+Q$, but when acceleration ends ($|\dot H/H^2|\sim 1$) and $\epsilon\simeq 1+Q$, $\rho_{rad}\simeq Q/(2(1+Q))$, such that in strong dissipation regime with $3H\leq \mathcal Y$, the radiation becomes relevant after some times.

\section{Inhomogeneous fluids and warm inflation}

Accelerated cosmology is the result of repulsive gravity in our universe. Standard matter and radiation cannot play this role. To describe inflation, an (effective) fluid which violates the Strong energy condition (the ratio between pressure and energy density must be smaller than $-1/3$) has to be introduced in the theory. This effective fluid can be given by a scalar field with potential, can be the result of some modifications to the gravitational action of Einstein's theory or can be merely an exotic ``dark'' fluid with an (effective) negative pressure. Generally speaking, it is clear that, in order to exit from inflation, the fluid must have an inhomogeneous Equation of State (EoS) parameter,
\begin{equation}
\omega(\rho)=\frac{p}{\rho}\,,
\end{equation}
depending on the energy density. Here, $p\,,\rho$ are the pressure and the energy density of the fluid. Our aim is to study fluid cosmology in the warm inflation scenario, generalizing the results of scalar field inflation. 
The Friedmann equations are
\begin{equation}
\frac{3 
H^2}{\kappa^2}=\rho_{rad}+\rho\,,\quad-\frac{1}{\kappa^2}\left(2\dot 
H+3 H^2\right)=p_{rad}+p\,,\label{EOMs}
\end{equation}
where the radiation contribute has been also considered,
and the conservation law leads to
\begin{eqnarray}
\dot\rho+3H\rho(1+\omega(\rho))&=&-\mathcal Y f(\rho)\,,\nonumber\\
\dot\rho_{rad}+4H\rho_{rad}&=&\mathcal Y f(\rho)\,,\quad \quad 0<\mathcal Y\,, f(\rho)\,.
\label{conslaw}
\end{eqnarray}
Here, $f(\rho)$ is a general (positive) function of the energy density of the fluid with the dimension $[f(\rho)]=[\rho]$, and therefore $[\mathcal Y]=[1/\kappa]$ is a friction coefficient and it is assumed to be positive to produce positive energy density for radiation. Up to now, the only requirement is that $3H\rho(1+\omega(\rho))\sim \mathcal Y f(\rho)$ during inflation, $3H\rho | (1+\omega(\rho))|< \mathcal Y f(\rho)$ corresponding to strong dissipation regime, but $\mathcal Y f(\rho)\ll H\rho$ to avoid radiation contribute to the dynamics.\\
\\
The (quasi) de Sitter solution of inflation evolves with the (positive) Hubble flow functions
\begin{equation}
\epsilon_1=-\frac{\dot H}{H^2}\,,\quad 
\epsilon_2=-\frac{2\dot H}{H^2}+\frac{\ddot H}{H\dot H}\equiv \frac{\dot\epsilon_1}{H\epsilon_1}\,,\label{Hflow}
\end{equation}
which have to remain small until the end of inflation, when acceleration finishes end $\epsilon_1\simeq 1$ (note that $\dot H<0$). By assuming $H^2\simeq \kappa^2\rho/3$, we get for these functions,
\begin{equation}
\epsilon_1=\frac{3(\omega(\rho)+1)}{2}+\frac{\mathcal Y}{2H}\left(\frac{f(\rho)}{\rho}\right)\,,\quad
\epsilon_2=\frac{3H\dot\omega(\rho)+\mathcal Y\left(\dot f(\rho)/\rho\right)}{3H^2(1+\omega(\rho))+\mathcal Y H\left(f(\rho)/\rho\right)}+\frac{3\mathcal Y f(\rho)}{2H\rho}\,.
\label{Hf}
\end{equation}
When $\mathcal Y=0$ we recover the formulas for classical inflationary scanario, while by making use of the relations (\ref{relphi}) with the slow-roll approximation (\ref{sa}) we find (\ref{epsilonetaphi}). 

One may also introduce the $e$-folds number left to the end of inflation,
\begin{equation}
N=\ln\left[\frac{a(t_\text{f})}{a(t)}\right]\,,\label{N}
\end{equation}
where $a(t_\text{i})\,,a(t_\text{f})$ are the scale factor at the beginning and at the end of inflation, respectively, $t_\text{i,f}$ being the related times.
The total amount of inflation is measured by 
\begin{equation}
\mathcal N\equiv \ln \left(\frac{a_\mathrm{f}(t_\text{f})}{a_\mathrm{i} (t_\text{i})}\right)=\int^{t_\text{f}}_{t_\text{i}} H(t)
dt\,,\label{Nfolds}
\end{equation}
and in order to have termalization at the end of inflation we must require, according with the spectrum of CMB fluctuations,  $55<\mathcal N<65$. 
Thus, the spectral index $n_s$ and the tensor-to-scalar 
ratio $r$ read
\begin{equation}
n_s=1-2\epsilon_1-\epsilon_2
\,,\quad r=16\epsilon_1
\,,\label{index}
\end{equation}
where $\epsilon_{1,2}$ are evaluated at $N=\mathcal N$.
The last cosmological data coming by the Planck satellite~\cite{Planckdata} constrain these indexes as
$n_{\mathrm{s}} = 0.9603 \pm 0.0073\, (68\%\,\mathrm{CL})$ and 
$r < 0.11\, (95\%\,\mathrm{CL})$.
Inflation parameters for fluid were studied also in Ref.~\cite{ip1,ip2}.

In the next section we will furnish several examples of fluid models realizing warm inflation in agreement with the Planck data.
\section{Viable fluid models for warm inflation}

A suitable Ansatz for the Equation of State parameter of fluid driving inflation in terms of the  $e$-folds left to the end of inflation ($0<N<\mathcal N\simeq 60$) may be given by~\cite{muk1,muk2},
\begin{equation}
1+\omega(\rho)=\frac{\beta}{(N+1)^\alpha}\,,\quad 0<\alpha\,,
\label{OmAn}
\end{equation}
where $\beta$ is a number on the order of the unit and, if $\mathcal Y\neq 0$ in (\ref{conslaw}), it can be either positive or negative.
In this way, the inflation is realized when $N=\mathcal N\simeq 60$ and $\omega$ is close to minus one (since the friction term $\mathcal Y f(\rho)$ is small, this condition is necessary to have a quasi-de Sitter universe). In the absence of dissipation ($\mathcal Y=0$), inflation must take place in the quintessence region with $0<\beta$, but here, thanks to the contribute of $\mathcal Y f(\rho)$ in the equations of motion, other scenarios are possible: if $0<\beta$, it is clear that the (positive) friction term must be extremelly small to have $\epsilon_1\,,\epsilon_2\ll 1$ in (\ref{Hf}); on the other hand, in the strong dissipation regime with $|3H(1+\omega(\rho))|<\mathcal Y f(\rho)$, $\dot H$ is negative and inflation ends at $\dot H\simeq -H^2$ even if $\beta$ is negative and the Equation of State parameter of fluid is smaller than minus one. 
Moreover, while when $\mathcal Y=0$ the quantity $(1+\omega(\rho))$ cannot vanish to have a graceful exit from inflation,  here we may consider also the case $\omega=-1$, since to the effective pressure of the fluid we get a contribution from the coupling with radiation.   

We can analyze some cases of inhomogeneous fluids for inflation under the Ansatz (\ref{OmAn}): from the first equation in (\ref{conslaw}), when the fluid is dominant and $H^2=\kappa^2\rho/3$, one finds the equation
\begin{equation}
-\frac{d\rho}{dN}+3\rho
\frac{\beta}{(N+1)^\alpha}=-\mathcal Y \left(\frac{f(\rho)}{\rho}\right)\sqrt{\frac{3}{\kappa^2}}\,,
\end{equation}
whose solution is the fluid energy density as a function of $N$ and it immediatly follows the explicit form of $\omega(\rho)$. 
Thus, the analysis of the Hubble flow functions permits to verify the viability of the model. In the next subsections, we will consider different forms of friction function  
for inhomogeneous fluid coupled with radiation. The special case $\omega=-1$ will be also discussed.

\subsection{Friction term with $f(\rho)=\pm\left(\rho/\rho_0\right)^{3/2}(1+\omega(\rho))/\kappa^4$\label{firstex}}

As a first example of inhomogeneous fluid for warm inflation, we look for the following simple form of friction function,
\begin{equation}
f(\rho)=\pm\frac{1}{\kappa^4}\left(\frac{\rho}{\rho_0}\right)^{3/2}\left(1+\omega(\rho)\right)\,,
\end{equation}
where $\rho_0$ is the energy density at the beginning of inflation, the Planck Mass in $\kappa^{1/4}$ has been introduced for dimensional reasons and the sign plus/minus depends on the sign of $(1+\omega)$ in order to have $0<f(\rho)$. Thus, the dissipative term in (\ref{conslaw}) is small during de Sitter expansion.

Let us reconstruct the explicit form of $\omega(\rho)$ and $f(\rho)$ by starting from
the Ansatz (\ref{OmAn}) with $\beta\neq 0$. From the first equation in (\ref{conslaw}) one has
\begin{equation}
\rho=\rho_\text{f}(N+1)^{3\beta\left(1+Q\right)}\,,\quad \alpha=1\,,
\end{equation}
\begin{equation}
\rho=\rho_0 \text{e}^{-\frac{3\beta}{(\alpha-1)(N+1)^{\alpha-1}}\left(1+Q\right)}\,, \quad\alpha\neq 1\,,
\end{equation}
where in the first case $\rho_\text{f}$ is the value of the energy density of the fluid at the end of inflation such that $\rho_\text{f}\ll\rho_0$ when $N=0$, while in the second case $\rho\simeq\rho_0$ when $1\ll N$ at the beginning of inflation. 
It is immediatly clear that $0<\beta(1+Q)$ if $1\leq\alpha$.

The number
\begin{equation}
Q=\pm\frac{\mathcal Y}{\sqrt{3}\kappa^5\rho_0^{3/2}}\,,
\end{equation}
encodes the contribute of the friction coefficient. 
Since the strong dissipation regime corresponds to $1<|Q|$ and $Q$ is a constant, in this model this phase cannot strongly appears (otherwise, $\rho$ does not remain constant during inflation). 
We also remember that $0<Q$ when $0<\beta$ and $Q<0$ when $\beta<0$. 

The explicit forms of $\omega(\rho)$ and $f(\rho)$ which describe the inhomogeneous fluid and its coupling with radiation are given by
 \begin{equation}
\omega(\rho)=-1+\beta\left(\frac{\rho_\text{f}}{\rho}\right)^{\frac{1}{3\beta(1+Q)}}\,,
\quad f(\rho)=\pm\frac{\beta}{\kappa^4}\left(\frac{\rho_\text{f}}{\rho_0}\right)^{3/2}\left(\frac{\rho_\text{f}}{\rho}\right)^{\frac{2-9\beta(1+Q)}{6\beta(1+Q)}}\,,\quad
\alpha=1\,,
\end{equation}
\begin{equation}
\hspace{-0.5cm}\omega(\rho)=-1+\beta\left[\frac{(\alpha-1)}{3\beta(1+Q)}\log\left[\frac{\rho_0}{\rho}\right]\right]^{\frac{\alpha}{\alpha-1}}\,,\quad
f(\rho)=\pm\frac{\beta}{\kappa^4}\left(\frac{\rho}{\rho_0}\right)^{3/2}
\left[\frac{(\alpha-1)}{3\beta(1+Q)}\log\left[\frac{\rho_0}{\rho}\right]\right]^{\frac{\alpha}{\alpha-1}}\,,\quad
\alpha\neq 1\,.
\end{equation}
Let us see for which values of $\alpha\,,\beta$ we have viable inflation.
The Hubble flow functions in (\ref{Hf}) rewritten in terms of $N$ read in our case
\begin{eqnarray}
\epsilon_1\equiv\frac{3}{2}(1+\omega(\rho))\left(1+Q\right)&=&\frac{3}{2}\frac{\beta}{(1+N)^{\alpha}}\left(1+Q\right)\,,\nonumber\\
\epsilon_2\equiv-\frac{1}{\epsilon_1}\frac{d\epsilon_1}{d N}&=&\frac{\alpha}{(N+1)}\,.
\end{eqnarray}
By evaluating this functions at $N=\mathcal N$ with $\mathcal N\simeq 60$, we get the spectral index and the tensor-to-scalar ratio in (\ref{index}),
\begin{equation}
n_s=1-\left[\frac{3\beta(1+Q)+\alpha(\mathcal N+1)^{\alpha-1}}{(\mathcal N+1)^{\alpha}}\right]\,,
\quad r=\frac{24\beta(1+Q)}{(\mathcal N+1)^{\alpha}}\,.
\end{equation}
For $\alpha=1$ one has
\begin{equation}
(1-n_s)=\frac{3\beta(1+Q)+1}{(\mathcal N+1)}\,,\quad
r=\frac{24\beta(1+Q)(1-n_s)}{3\beta(1+Q)+1}\,.
\end{equation}
To satisfy the Planck data when $\mathcal N=60$, it must be
$\beta\simeq 1/(3(1+Q))$. 
If $0<\beta$ and $0<Q$, in order to have viable inflation, $0<\beta<1/3$. On the other side, if $\beta<0$ and $Q<-1$, inflation is viable for $-1/3<\beta<0$, namely, as we noted in the introduction to this section, acceleration can finish even if $\omega$ remains smaller than minus one.
In both of the cases, the Equation of State parameter of the fluid and the coupling with radiation are given by
\begin{equation}
\omega(\rho)=-1+\frac{1}{3(1+Q)}\left(\frac{\rho_\text{f}}{\rho}\right)\,,\quad
f(\rho)=\pm\frac{1}{3(1+Q)\kappa^4}\left(\frac{\rho_\text{f}}{\rho_0}\right)^{3/2}\left(\frac{\rho}{\rho_\text{f}}\right)^{1/2}\,,\label{ex1}
\end{equation}
where $\rho_\text{f}$ is a constant corresponding to the energy density of the fluid at the end of inflation and $\rho_0$ is a constant indicating the energy density of the fluid at the beginning of inflation. We note that, during the fluid dominance expansion,  $\mathcal Y f(\rho)\ll H\rho$.

For $1<\alpha$ one gets
\begin{equation}
(1-n_s)\simeq\frac{\alpha}{(\mathcal N+1)}\,,\quad
r\simeq\frac{24\beta(1+Q)(1-n_s)^\alpha}{\alpha^\alpha}\,,
\end{equation}
and for $0<\alpha<1$ we obtain,
\begin{equation}
(1-n_s)\simeq\frac{3\beta(1+Q)}{(\mathcal N+1)^{\alpha}}\,,\quad
r\simeq\frac{24\beta(n_s-1)}{3\beta(1+Q)}\,.
\end{equation}
In this cases, the only possibility to satisfy the Planck data is given by $\alpha=2$ ($2<\alpha$ renders to small the spectral index and $0<\alpha<1$ too large the tensor-to-scalar ratio). Since $r$ is extremely small, every value of $\beta$ (on the order of the unit) is allowed for $\alpha=2$. Also in this case, when $\beta<0$ and $Q<-1$, we obtain inflation with graceful exit even if $\omega<-1$.
The explicit forms of the EoS parameter and the friction function read
\begin{equation}
\omega(\rho)=-1+\frac{1}{9\beta(1+Q)^2}\left[\log\left[\frac{\rho_0}{\rho}\right]\right]^{2}\,,\quad
f(\rho)=\pm\frac{1}{\kappa^4}\left(\frac{\rho}{\rho_0}\right)^{3/2}
\frac{1}{9\beta(1+Q)^2}\left[\log\left[\frac{\rho_0}{\rho}\right]\right]^{2}\,,\label{ex2}
\end{equation}
where $\rho_0$ is the boundary value of the energy density of the fluid at the beginning of inflation. Also in this case, $\mathcal Y f(\rho)\ll H\rho$ during accelerated expansion.

In this examples, the ratio between energy density of radiation and inhomogeneous fluid (\ref{ex1}) or (\ref{ex2}) is derived by the solution of the second equation in (\ref{conslaw}), namely
\begin{equation}
\frac{\rho_{rad}}{\rho}\simeq\frac{\epsilon_1 Q}{2(1+Q)}\,,
\end{equation}
such that, when $\epsilon_1 \ll 1$, one has $\dot\rho_{rad}\simeq 0$. During inflation the energy density of radiation is almost a constant and negligible, but at the end of it, when $\epsilon_1\simeq 1$, it grows up and the radiation era takes place, completing the warm inflationary scenario. In the strong dissipation regime with $Q\leq-1$, namely $\omega(\rho)<-1$, radiation can be dominant respect to the fluid at the end of early-time acceleration. On the other hand,
if $0<Q$ and $-1<\omega(\rho)$, in a first moment the decelerated expansion is driven by fluid and radiation together, and only when $2<\epsilon_1$ the radiation component becomes dominant (like in scalar field representation): such a process is well realized in strong dissipation regime with $1\leq Q$.

In the example (\ref{ex2}) the value of the friction function at the end of inflation is larger than its value at the onset of the early-time acceleration, namely
$0=|f(\rho_0)|< |f(\rho_\text{f})|$, leading to a fluid EoS parameter far from minus one. On the other hand,
in the example (\ref{ex1}), the magnitude of the friction function decreases during inflation and tends to vanish. As a consequence, the behaviour of the fluid becomes the one of a perfect fluid with EoS parameter $\omega=1/(3(1+Q))$, and we may speculate that, if $1\ll |Q|$ (it means, $|\beta|\ll 1$), a dark energy fluid could emerge from Friedmann universe.

\subsection{Friction term with $f(\rho)=\pm(\rho_0-\rho)(1+\omega(\rho))\sqrt{\rho/\rho_0}$\label{secondex}}

Let us consider now an other suitable friction function whose form is given by 
\begin{equation}
f(\rho)=\pm\left(\rho_0-\rho\right)\left(1+\omega(\rho)\right)\sqrt{\frac{\rho}{\rho_0}}\,,
\end{equation}
where $\rho_0$, as usually, is the energy density of the fluid at the beginning of inflation and the sign plus/minus corresponds to the sign of $(1+\omega)$ such that $0<f(\rho)$, like in the preceeding example. The behaviour of this coupling function is similar to the one in (\ref{consrad}) for scalar field warm inflation.

The form of the fluid energy density from
the Ansatz (\ref{OmAn}) with $\beta\neq 0$ is derived as
\begin{equation}
\rho=\rho_1(N+1)^{3\beta\left(1-Q\right)}-\frac{Q\rho_0}{(1-Q)}\,,\quad \alpha=1\,,
\end{equation}
\begin{equation}
\rho=\frac{\rho_0}{(1-Q)} \left(\text{e}^{-\frac{3\beta}{(\alpha-1)(N+1)^{\alpha-1}}\left(1-Q\right)}-Q\right)\,, \quad\alpha\neq 1\,,
\end{equation}
where in the first case $\rho_1=\rho_\text{f}+Q\rho_0/(1-Q)$, where 
$\rho_\text{f}$ is the value of the energy density of the fluid at the end of inflation (note that $\beta(1-Q)$ can be positive or negative when $1\leq\alpha$). 
Again, the number
\begin{equation}
Q=\pm\frac{\mathcal Y}{\sqrt{3}\kappa\sqrt{\rho_0}}\,,
\end{equation}
indicates the contribution of the friction term to de solution. The strong dissipation regime corresponds to $1<|Q\left(\rho_0/\rho-1\right)|$ and at the end of inflation the model strongly is in this phase.

The explicit forms of $\omega(\rho)$ and $f(\rho)$ read
 \begin{eqnarray}
&&\omega(\rho)=-1+\beta\left(\frac{\rho_1}{\rho+Q\rho_0/(1-Q)}\right)^{\frac{1}{3\beta(1-Q)}}\,,\nonumber\\&&
\hspace{3cm}f(\rho)=\pm\left(\rho_0-\rho\right)\beta\left(\frac{\rho_1}{\rho+Q\rho_0/(1-Q)}\right)^{\frac{1}{3\beta(1-Q)}}\sqrt{\frac{\rho}{\rho_0}}\,,\quad
\alpha=1\,,\label{exex1}
\end{eqnarray}
\begin{eqnarray}
&&\omega(\rho)=-1+\beta\left[\frac{(\alpha-1)}{3\beta(1-Q)}\log\left[\frac{\rho_0}{\rho(1-Q)+Q\rho_0}\right]\right]^{\frac{\alpha}{\alpha-1}}\,,\nonumber\\&&
\hspace{1cm}
f(\rho)=\pm\left(\rho_0-\rho\right)\beta\left[\frac{(\alpha-1)}{3\beta(1-Q)}\log\left[\frac{\rho_0}{\rho(1-Q)+Q\rho_0}\right]\right]^{\frac{\alpha}{\alpha-1}}\sqrt{\frac{\rho}{\rho_0}}\,,\quad
\alpha\neq 1\,.\label{exex2}
\end{eqnarray}
Let us analyze the values of 
$\alpha\,,\beta$ which lead to realistic warm inflation.
The Hubble flow functions in (\ref{Hf}) evaluated for $1\ll N$, namely when $\rho\simeq\rho_0$, read 
\begin{eqnarray}
\epsilon_1&\equiv&\frac{3}{2}(1+\omega(\rho))\left[1+Q\left(\frac{\rho_0}{\rho}-1\right)\right]\simeq\frac{3}{2}\frac{\beta}{(1+N)^{\alpha}}\,,\nonumber\\
\epsilon_2&\equiv&-\frac{1}{\epsilon_1}\frac{d\epsilon_1}{d N}\simeq\frac{\alpha}{(N+1)}+\frac{3\beta Q}{(N+1)^\alpha}\,.
\end{eqnarray}
The spectral index and the tensor-to-scalar ratio in (\ref{index}) for $N=\mathcal N\simeq 60$ are derived as
\begin{equation}
n_s\simeq 1-\frac{3\beta(1+Q)+\alpha(\mathcal N+1)^{\alpha-1}}{(1+\mathcal N)^\alpha}\,,\quad
\epsilon_2\simeq\frac{24\beta}{(N+1)^\alpha}\,.
\end{equation}
Thus, the form of the results is very similar to the case analyzed in \S~\ref{firstex}, but we must remember that the definition of $Q$ changes. Also here, in order to satisfy the Planck data, we have only two possibilities: $\alpha=1$ or $\alpha=2$. When 
$\alpha=1$ we must choose $\beta(1+Q)\simeq 1/3$, while when $\alpha=2$ every value of $\beta$ (on the order of the unit) is allowed. Also in this case, we may have inflation with graceful exit even for $\omega(\rho)<-1$.

The ratio between energy density of radiation and inhomogeneous fluid is given by the solution of the second equation in (\ref{conslaw}), 
\begin{equation}
\frac{\rho_{rad}}{\rho}\simeq\frac{\epsilon_1 Q(\rho_0/\rho-1)}{2(1+Q(\rho_0/\rho-1))}\,,
\end{equation}
such that, when $\epsilon_1 \ll 1$ and $\rho\simeq\rho_0$, the radiation contribute is small, but at the end of inflation, when $\epsilon_1$ is on the order of the unit and $\rho\ll\rho_0$ such that $1\ll |Q(\rho_0/\rho-1)|$ in very strong dissipation regime, it grows up and radiation era takes place. 

In the fluids analyzed in this subsection, the friction function grows up during inflation and in general $0=|f(\rho_0)|<|f(\rho_\text{f})|$.
In the case of example (\ref{exex1}), its value at the end of the early-time acceleration results to be $f(\rho_\text{f})\simeq\pm\sqrt{\rho_0\rho_\text{f}}$ with $\omega(\rho_\text{f})=-1+\beta$. More interesting is the case of the example (\ref{exex2}) with $\alpha=2$: for large values of $\beta$ the friction function remains small at the end of inflation and may lead to the appearance  of a phantom/quintessence-like fluid with EoS parameter close to minus one in the Friedmann universe.

\subsection{Friction term with $\omega=-1$ and $f(\rho)=(\rho/\rho_0)^\alpha(\rho_f/\rho)^\gamma/\kappa^4$\label{thirdex}}

As the last example, we introduce a fluid with constant EoS parameter $\omega=-1$ and coupled with matter as in (\ref{conslaw}) throught the following friction function,
\begin{equation}
f(\rho)=\frac{1}{\kappa^4}\left(\frac{\rho}{\rho_0}\right)^\alpha\left(\frac{\rho_\text{f}}{\rho}\right)^\gamma,\quad 0<\alpha\,,\gamma\,,
\end{equation}
where $\rho_0\,,\rho_\text{f}$ are the energy density at the beginning and at the end of inflation. The Mass Planck in $\kappa^4$ has been introduced for dimensional reason.  
We note that $f(\rho)$ remains small during inflation when $\rho_{\text{f}}\ll\rho<\rho_0$.

Thus, the first equation in (\ref{conslaw}) with the first equation in (\ref{EOMs}) when the fluid is dominant can be rewritten in terms of $N$ as
\begin{equation}
\frac{d\rho}{dN}=\mathcal Y f(\rho)\sqrt{\frac{3}{\kappa^2\rho}}\,.
\end{equation}
The solution of the above equation reads
\begin{equation}
\rho=\left[\left(\tilde\gamma+\frac{3}{2}\right)\left(3 Q N+c_1\right)
\right]^{1/(\tilde\gamma+3/2)}\,,\quad \tilde\gamma=(\gamma-\alpha)\,,
\end{equation}
where $c_1$ is a constant such that $\rho_\text{f}=\left[(\tilde\gamma+3/2)c_1\right]^{1/(\tilde\gamma+3/2)}$, and 
\begin{equation}
Q=\frac{\mathcal Y\rho_\text{f}^\gamma}{\sqrt{3}\kappa^5\rho_0^\alpha}\,.
\end{equation}
In this model, the strong dissipation regime strongly appears at the end of inflation, when $\rho_\text{f}\simeq\rho\ll\rho_0$.

The Hubble flow functions in (\ref{Hf}) are given by
\begin{eqnarray}
\epsilon_1&\equiv&
\frac{\mathcal Y}{2H}\left(\frac{f(\rho)}{\rho}\right)=
\frac{3Q}{(3+2\tilde\gamma)\left(c_1+3NQ\right)}
\,,\nonumber\\
\epsilon_2&\equiv&-\frac{1}{\epsilon_1}\frac{d\epsilon_1}{d N}=
\frac{3Q}{c_1+3NQ}\,.
\end{eqnarray}
When $1\ll N$,  we observe that
$\epsilon_1\,,\epsilon_2\ll 1$, and
when $N$ decreases, the Hubble flow functions increase. The spectral index and the tensor-to-scalar ratio in (\ref{index}) are
derived as
\begin{equation}
n_s=  1-
\frac{3Q(5+2\tilde\gamma)}{(3+2\tilde\gamma)(c_1+3\mathcal NQ)}
\,,\quad
r=\frac{48Q}{(3+2\tilde\gamma)(c_1+3\mathcal N Q)}\,.
\end{equation}
Thus, in order to satisfy the last Planck data when $(c_1+3\mathcal N Q)\simeq 3\mathcal N Q$ and $\mathcal N\simeq 60$, we need $\tilde\gamma\simeq -0.78$, namely $\gamma<\alpha$. 
We should note that in this case the tensor-to-scalar ratio $r$ results to be slightly larger than the Planck constraint, but, since its value is still a debated question, the result may be interesting.

Now the ratio between energy density of radiation and energy density of fluid is
\begin{equation}
\frac{\rho_{rad}}{\rho}=\frac{\epsilon_1}{2}\,,
\end{equation}
and remains small and almost a constant during early-time acceleration induced by fluid, but it grows up at the end and radiation era takes place. 

We would like to note that at the end of inflation, since the friction function tends to vanish, the fluid here analyzed becomes in fact a dark energy fluid.

%%%%%%%%%%%%%%%%%%%%%%%%
%%%  Acknowledgments
%%%%%%%%%%%%%%%%%%%%%%%%
%\section*{Acknowledgments}

\section{Conclusions}

In this paper, we applied the warm inflationary scenario to fluid cosmology. An inhomogeneous fluid coupled with radiation drives the early-time acceleration, and during this period the energy density of radiation is almost a constant and leads to a radiation dominated expansion after the end of inflation. 
In the last decade, warm inflation has been well studied especially in the framework of scalar field theories, where the interaction between the kinetic energy of inflaton and radiation fields is given by a friction coefficient $\mathcal Y$, and the radiation dominated era takes place after inflation in the strong dissipation regime, when $3H<\mathcal Y$. We have considered several models of inhomogeneous fluids realizing warm inflation in the presence of different couplings with radiation throught some functions $\mathcal Y f(\rho)$. The dissipation regime appears in different ways: in the specific, in the example of \S~\ref{firstex} the ratio $|\mathcal Y f(\rho)/(3H (1+\omega))|$ is a constant, while in the examples of \S~\ref{secondex} and \S~\ref{thirdex} the strong dissipation regime takes place at the end of inflation. In the last subsection a fluid with constant EoS paramter $\omega=-1$ has been considered and, thanks to the coupling with radiation, we can obtain a graceful exit from inflation. For all the models under investigation we analyzed and discussed the confidence with the Planck data and the emerging of radiation epoch at the end of inflation.

Some detailed works on fluid cosmology can be found also in Refs.~\cite{Alessia, Alessia2, Brevik, davood1, davood2, davood3}. Other works on fluid cosmology applied to inflation are in Refs.~\cite{fi1,fi2}.

\end{document}